\documentclass[twocolumn]{aastex701}

\newcommand{\cii}{[C\,{\sc ii}]}

\usepackage{amsmath, amssymb, graphicx, xcolor, amsbsy,amsfonts}

\begin{document}

\title{High-Resolution ALMA Imaging for a Gravitationally-lensed Quasar at $z=6.5$:\\ Constraining the AGN Contribution to Galactic-Scale Dust Heating}

\author[0000-0002-5367-8021]{Minghao Yue}
\email{myue@mit.edu}
\affiliation{Steward Observatory, University of Arizona, 933 North Cherry Ave., Tucson, AZ 85721, USA}

\author[0000-0003-3310-0131]{Xiaohui Fan}
\affiliation{Steward Observatory, University of Arizona, 933 North Cherry Avenue, Tucson, AZ 85721, USA}
\email{}

\author[0000-0003-2895-6218]{Anna-Christina Eilers}
\email{eilers@mit.edu}
\affiliation{MIT Kavli Institute for Astrophysics and Space Research, 77 Massachusetts Ave., Cambridge, MA 02139, USA}

\author[0000-0001-5287-4242]{Jinyi Yang}
\affiliation{Department of Astronomy, University of Michigan, 1085 S. University Ave., Ann Arbor, MI 48109, USA}
\email{}

\author[0000-0002-7633-431X]{Feige Wang}
\affiliation{Department of Astronomy, University of Michigan, 1085 S. University Ave., Ann Arbor, MI 48109, USA}
\email{}

\author[0000-0002-6221-1829]{Jianwei Lyu}
\affiliation{Steward Observatory, University of Arizona, 933 North Cherry Avenue, Tucson, AZ 85721, USA}
\email{}

\author[0000-0002-8987-7401]{James W. Nightingale}
\affiliation{School of Mathematics, Statistics and Physics, Newcastle University, Herschel Building, Newcastle-upon-Tyne, NE1 7RU, UK}
\email{}

\author[0000-0001-6047-8469]{Ann Zabludoff}
\affiliation{Steward Observatory, University of Arizona, 933 North Cherry Avenue, Tucson, AZ 85721, USA}
\email{}







\author[0000-0003-4956-5742]{Ran Wang}
\affiliation{Department of Astronomy, School of Physics, Peking University, Beijing 100871, P. R. China}
\affiliation{Kavli Institute for Astronomy and Astrophysics, Peking University, Beijing 100871, P. R. China}
\email{}

\begin{abstract}

We present high-resolution (beam size $0\farcs076\times0\farcs040$) Atacama Large Millimeter/submillimeter Array (ALMA)  observations of the far-infrared $(\lambda_\text{rest}=162.7\mu\rm{m})$ dust continuum of J0439+1634, a gravitationally lensed quasar at $z=6.52$. We perform pixelated lens modeling for the visibility data, finding that J0439+1634 is well-described by a singular isothermal ellipsoid plus an external shear lensing model. The best-fit lensing potential exhibits a naked-cusp configuration, confirming the finding in \cite{fan19}. The reconstructed source plane continuum emission shows a compact bright core, with size $\lesssim200$ pc and peak brightness $\sim0.6 \text{ Jy arcsec}^{-2}$. The total continuum flux at 245 GHz is {$3.36\pm0.02$ mJy}. The flux magnification is {$4.63\pm0.03$}, indicating an average source-plane resolution of $0\farcs019$ (equivalent to 104 pc). 
The spatial resolution around the supermassive black hole reaches $\sim36$ pc.
Leveraging the exceptional source-plane resolution, we build a radiative transfer model to describe the observed dust emission profile. The best-fit model indicates that heated dust from the active galactic nucleus (AGN)  dominates the sub-millimeter emission at $r\lesssim100$ pc and that star-heated dust dominates the outer region of the host galaxy. AGN heating contributes {$\sim13\%$} to the observed sub-mm flux. Therefore, previous far-infrared-based star formation rate measurements for most high-redshift quasars are likely mildly overestimated.


\end{abstract}

\keywords{}


\section{Introduction} 

The discoveries of luminous quasars at $z\gtrsim6.5$ \citep[e.g.,][]{banados18,matsuoka19,wang19a,wang21,yang19a,yang20,fan23} indicate that supermassive black holes (SMBHs) already exist within the first Gyr of cosmic history. These early SMBHs already show strong coevolution with their host galaxies \citep[e.g.,][]{yue24,marshall25,liu25, onoue25,zhu25}. 
Thus far, most of our knowledge about high-redshift quasar host galaxies has been from sub-millimeter (sub-mm) observations, where quasar host galaxies are much more luminous than their central active galactic nuclei (AGN) \citep[e.g.,][]{lyu22}. In particular, previous observations by the Atacama Large Millimeter/submillimeter Array (ALMA) have found that $z\gtrsim6$ quasar host galaxies are luminous in rest-frame far infrared (FIR), indicating high star formation rates (SFRs) of $\sim10^2-10^3M_\odot \text{yr}^{-1}$ \citep[e.g.,][]{decarli18,izumi19,Mazzucchelli25}. These quasar hosts have typical sizes of $\sim2-4$ kpc, showing diverse kinematics and morphologies \citep[e.g.,][]{venemans20,pensabene20,neeleman21,wang24}.

Most previous ALMA observations for $z\gtrsim6$ quasars have spatial resolutions $\gtrsim1$ kpc. To date, only very few $z>6$ quasars have ALMA images with resolution $<500$ pc \citep[][]{venemans19,walter22,meyer23}. 
The small-scale structures in quasar host galaxies might encode key information about AGN-galaxy coevolution in these systems \citep[][]{venemans19}. Therefore, it is important to expand the sample of $z>6$ quasars with high-resolution ALMA observations.

Gravitational lensing is a powerful tool that significantly enhances the spatial resolution of observations, providing detailed insights into the small-scale structures of distant objects. To this end, J043947.08+163415.7 ($z=6.52$; hereafter J0439+1634) offers an ideal case study. J0439+1634 is the only strongly-lensed $z>5$ luminous quasar known to date \citep[e.g.,][]{fan19,yang19,yue21}.
With the help of lensing magnification, we can probe the detailed structures in J0439+1634
that are not accessible for other high-redshift quasars. 

Here we present the high-resolution ALMA observation for J0439+1634. Leveraging the unparalleled angular resolution of ALMA, we constructed a new lensing model and performed pixelated reconstruction for the quasar host galaxy emission at sub-mm wavelengths. This paper focuses on the dust continuum emission of the quasar host galaxy, demonstrating how we can constrain the quasar's role in galactic-scale dust heating using high-resolution observations. We will further present the {\cii} line emission in the subsequent paper (Yue et al. in prep), where we will characterize the kinematics of the quasar host galaxy.

This paper is organized as follows. Section \ref{sec:data} describes the high-resolution ALMA observation of J0439+1634. Section \ref{sec:lensmodeling} presents lens modeling methods and results. Section \ref{sec:heating} discusses the AGN's contribution to dust heating at galactic scales. Section \ref{sec:conclusion} summarizes this paper. Throughout this paper, we assume a flat $\Lambda$CDM cosmology with $H_0=70\text{ km s}^{-1}\text{ Mpc}^{-1}$. 

\section{Data} \label{sec:data}

J0439+1634 was observed by ALMA under two configurations, namely C43-5 and C43-8. The program ID for the observation is 2018.1.00566.S (PI: Fan). The C43-5 observations were taken in 2018, with a total on-source time of 99 minutes. 
We use four spectral windows (SPWs) to sample the FIR continuum and the {\cii} emission line, which have central frequencies of 238.593 GHz, 236.718 GHz,
252.206 GHz, and 253.894 GHz, respectively.
All the SPWs have widths of 1.875 GHz.
The C43-5 configuration probes the large-scale $(\sim1'')$ emission of the quasar host galaxy and is crucial for measuring the total flux; the analysis for the 
 C43-5 observation has been presented in \citet{yue21}. 

The C43-8 observations were taken in 2019 and 2021, consisting of seven visits to the target. The total on-source exposure time for the C43-8 configuration is 337 minutes. The C43-8 configuration delivers a resolution of $0\farcs04$. We use the same spectral setup as that of the C43-5 observations.

\begin{figure}
    \centering
    \includegraphics[width=1\linewidth]{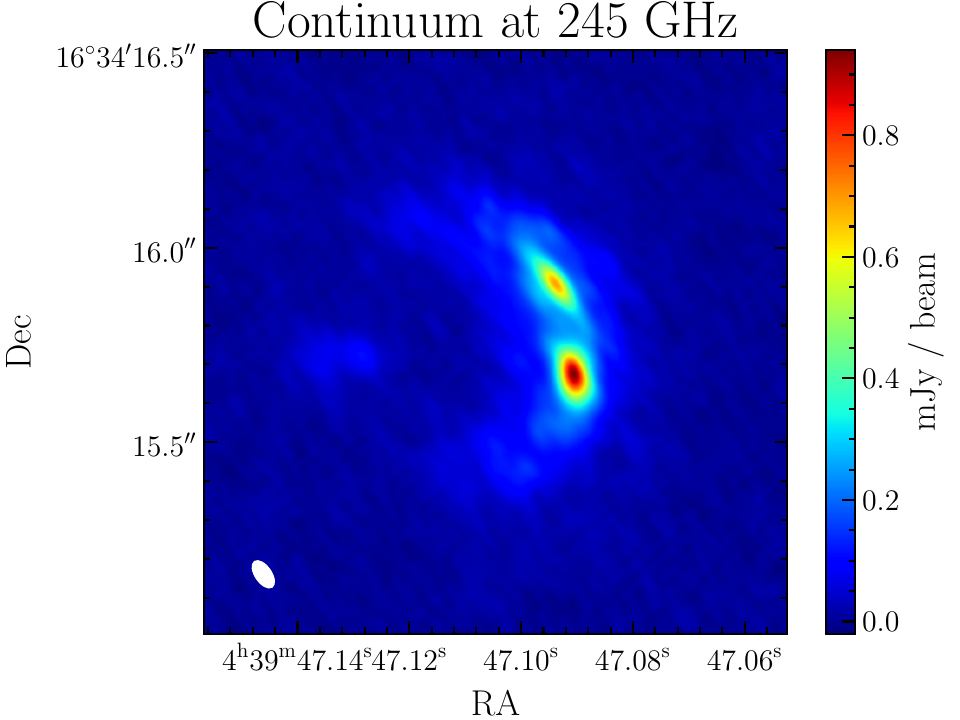}
    \caption{The clean image for the continuum emission of J0439+1634. The image is produced using the CASA task \texttt{tclean} using Briggs weighting (with robust $=0.5$). The beam size is $0\farcs078\times0\farcs040$, shown as the white ellipse at the lower leftcorner. The $1\sigma$ sensitivity is 0.02 mJy/beam. The two bright blobs are the lensed images of the quasar host galaxy's central region, and the diffused emission shows the lensed arc of the quasar host galaxy.}
    \label{fig:clean}
\end{figure}

We reduce the data using CASA \citep[][]{casa} version 6.5.4.9.
We first combine all the C43-5 and C43-8 observations into one measurement set, then obtain the 
continuum visibility by splitting out line-free channels using task \texttt{split}. Here we define line-free channels as channels with observed frequency $\nu_\text{obs}<252.4$ GHz or $\nu_\text{obs}>253.1$ GHz \citep{yue21}. 
We then run task \texttt{tclean}
using Briggs weighting (with robust$=0.5$) to obtain the clean image for the FIR continuum. 
Figure \ref{fig:clean} shows the resulting clean image. The beam size is $0\farcs078\times0\farcs040$. 

\section{Lens Modeling} \label{sec:lensmodeling}


\subsection{Pixelated Lensing Reconstruction} \label{sec:pixelated}

The goal of this study is to characterize the small-scale structures in J0439+1634's host galaxy down to $\lesssim100$ pc scale, which is likely irregular and cannot be well-described by analytical models (like S\'ersic profiles). Therefore, we perform pixelated lensing reconstruction for the dust continuum emission. The idea of pixelated lensing reconstruction has been described in, e.g., \citet{warren03,suyu06}, and was later applied to interferometric observations \citep[e.g.,][]{hezaveh16,litke19,maresca25}. Here we briefly describe the idea of pixelated reconstruction, and refer readers to the above-listed papers for more details.


A lensing model consists of a pixelated source $\vec{s}$ and a set of parameters $\eta$ describing the lensing operator $\textbf{L}(\eta)$. We can model the visibility of the lensing system as

\begin{equation} \label{eq:lens}
    \vec{m} = \textbf{D} \textbf{L}(\eta) \vec{s}
\end{equation}

where $\vec{s}$ is the source-plane emission, 
$\textbf{L}$ is the lensing operator that maps the source-plane emission to the image plane, 
$\textbf{D}$ represents the effect of the interferometer (including the primary beam and the Fourier transform),
 and $\vec{m}$ is the modeled UV-plane visibility.
 We note that $\textbf{D}$ and $\textbf{L}(\eta)$ are linear operators and can thus be represented by matrices.

Given the observed visibility data $\vec{d}$ and its covariance matrix $\boldsymbol{\Sigma}_d$, 
the log-likelihood of the data can be computed as

\begin{equation} \label{eq:loglike}
\log \mathcal{L}(\vec{d}|\eta)=-\frac{1}{2}[(\vec{d}-\vec{m})^\intercal \boldsymbol{\Sigma}_d(\vec{d}-\vec{m})]
\end{equation}


Previous studies \citep[e.g.,][]{hezaveh16,powell21,maresca25} have shown that, for a given lensing model $\textbf{L}(\eta)$, the best-fit source can be expressed as 
\begin{equation} \label{eq:source}
    \vec{s}=\textbf{A}^{-1}(\textbf{DL})^\intercal\boldsymbol{\Sigma}_d^{-1}\vec{d} 
\end{equation}
where $\textbf{A}=(\textbf{DL})^\intercal\boldsymbol{\Sigma}_d^{-1}(\textbf{DL}) + \lambda\textbf{R}^\intercal\textbf{R}$.
Here  $\textbf{R}$ is the regularization matrix to avoid overfitting, which punishes source models with large small-scale variations. The parameter $\lambda$ controls the strength of the regularization.
Plugging the derived $\vec{s}$ into Equation \ref{eq:lens} and \ref{eq:loglike} gives the log-likelihood of the observed data and allows evaluation of the posterior probability of lensing parameters $\eta$.

Practically, the pixelated source $\vec{s}$ is usually represented by a grid of points with a surface brightness assigned to each point. A widely used type of grid is a uniform grid on the image plane, which delivers finer spatial sampling on the source plane for higher-magnification regions. Another type of grid is the so-called brightness-adaptive grid, which has higher number densities of points in regions with higher surface brightness. The brightness-adaptive grid is suitable for sources with significant flux gradients in bright regions. We refer the readers to the \texttt{HowToLens} instructions\footnote{https://pyautolens.readthedocs.io/en/latest/howtolens/howtolens.html} for more information.

\subsection{Analysis for J0439+1634} \label{sec:analysis}

We use \texttt{PyAutoLens} \footnote{https://github.com/Jammy2211/PyAutoLens} \citep{pyautolens, Nightingale2015, Nightingale2018}, which can perform pixelated lensing reconstruction for interferometric observation. We use a singular isothermal ellipsoid (SIE) density profile plus an external shear to describe the lensing potential. This model has eight free parameters, namely the center position $(x, y)$, the Einstein radius, the position angle and the ellipticity of the SIE, the $x-$ and $y-$components of the external shear, and the regulation coefficient for the pixellated source. We list these parameters in Table \ref{tbl:model}.

\begin{deluxetable*}{cccc}
\tablenum{1}
\label{tbl:model}
\tablecaption{Parameters in the Lensing Model}
\tablewidth{0pt}
\tablehead{\colhead{Name} & \colhead{Meaning} & \colhead{Prior} & \colhead{Posterior (3-$\sigma$ limits)}}
\startdata
\hline
$x_\text{lens}~ ('')$ & The $x-$coordinate of the SIE center & Uniform at $[-0.559,-0.459]$  & $-0.5343^{+0.0018}_{-0.0038}$  \\
$y_\text{lens}~ ('')$ & The $y-$coordinate of the SIE center  & Uniform at $[-0.074,0.026]$ & $-0.0237_{-0.0004}^{+0.0005}$\\
$\theta_\text{E}~ ('')$ & The Einstein radius of the SIE & Uniform at $[0.33,0.43]$  & $0.4111_{-0.0036}^{+0.0070}$ \\
$e_1$ & The $y-$component of the SIE ellipticity & Uniform at $[-0.1198,0.0802]$ & $0.006_{-0.005}^{+0.003}$ \\
$e_2$ & The $x-$component of the SIE ellipticity & Uniform at $[0.4481, 0.6481]$ &  $0.644_{-0.011}^{+0.004}$ \\
$\gamma_1$ & The $y-$component of the external shear  & Uniform at $[-0.2,0.2]$  & $0.1850^{+0.0135}_{0.0099}$ \\ 
$\gamma_2$ & The $x-$component of the external shear & Uniform at $[-0.2,0.2]$ & $-0.0503^{+0.0081}_{-0.0087}$ \\ 
$\lambda$  & The regularization coefficient & Log-uniform at $[10^4,10^7]$ & $246573_{-28817}^{+20122}$
 \\\hline
\enddata
\tablecomments{The priors and posteriors correspond to the second \texttt{PyAutoLens} run. For $x_{\rm lens}$, $y_{\rm lens}$, $e_1$ and $e_2$, the mean value of the Uniform priors are the values from the first \texttt{PyAutoLens} run.}
\end{deluxetable*}

Following the strategy in \citet{maresca25}, we fit the lensing model in two steps. In the first step, we use a pixelated source corresponding to a $50\times50$ uniform grid covering the $2''\times2''$ area on the image plane. We assume flat priors for the model parameters, except for the regularization coefficient that have a log-uniform prior. This step aims at obtaining an approximated fit for the source emission. In the second step, we set up a brightness-adapted grid containing 600 pixels, where the number density of points on the image plane is proportional to the surface brightness as modeled in the first step. We then re-run the lens modeling to obtain the posterior distribution of the model parameters, with flat priors for the model parameters (again except for the regularization coefficient that have a log-uniform prior). In both steps, we use the constant split regularization provided by \texttt{PyAutoLens}. We run nested sampling using \texttt{nautilus} \citep{nautilus} to obtain the posterior distribution of the parameters. 

After obtaining the best-fit parameters, we use Equation \ref{eq:source} to determine the {best-fit} source plane emission, $\vec{s}$. {To characterize the uncertainty of the source plane emission, we randomly draw 100 sets of lensing parameters from the posterior distribution. For a given set of lensing parameters, the probability distribution of $\vec{s}$ is described as a multi-dimension Gaussian distribution, with a mean of $\vec{s}=\textbf{M}\vec{d}$ and a covariance matrix of $\boldsymbol{\Sigma}_s=\textbf{M}\boldsymbol{\Sigma}_d\textbf{M}^{\intercal}$ (where $\textbf{M}=\textbf{A}^{-1}(\textbf{DL})^\intercal\boldsymbol{\Sigma}_d^{-1}$, according to Equation \ref{eq:source}). We thus draw 100 random realizations for the source vector $\vec{s}$ for each set of lensing parameters, leading to a total number of 10,000 random sources. We evaluate the uncertainty of the source-plane measurements using these random sources.}

\begin{figure*}
    \centering
    \includegraphics[width=0.8\linewidth]{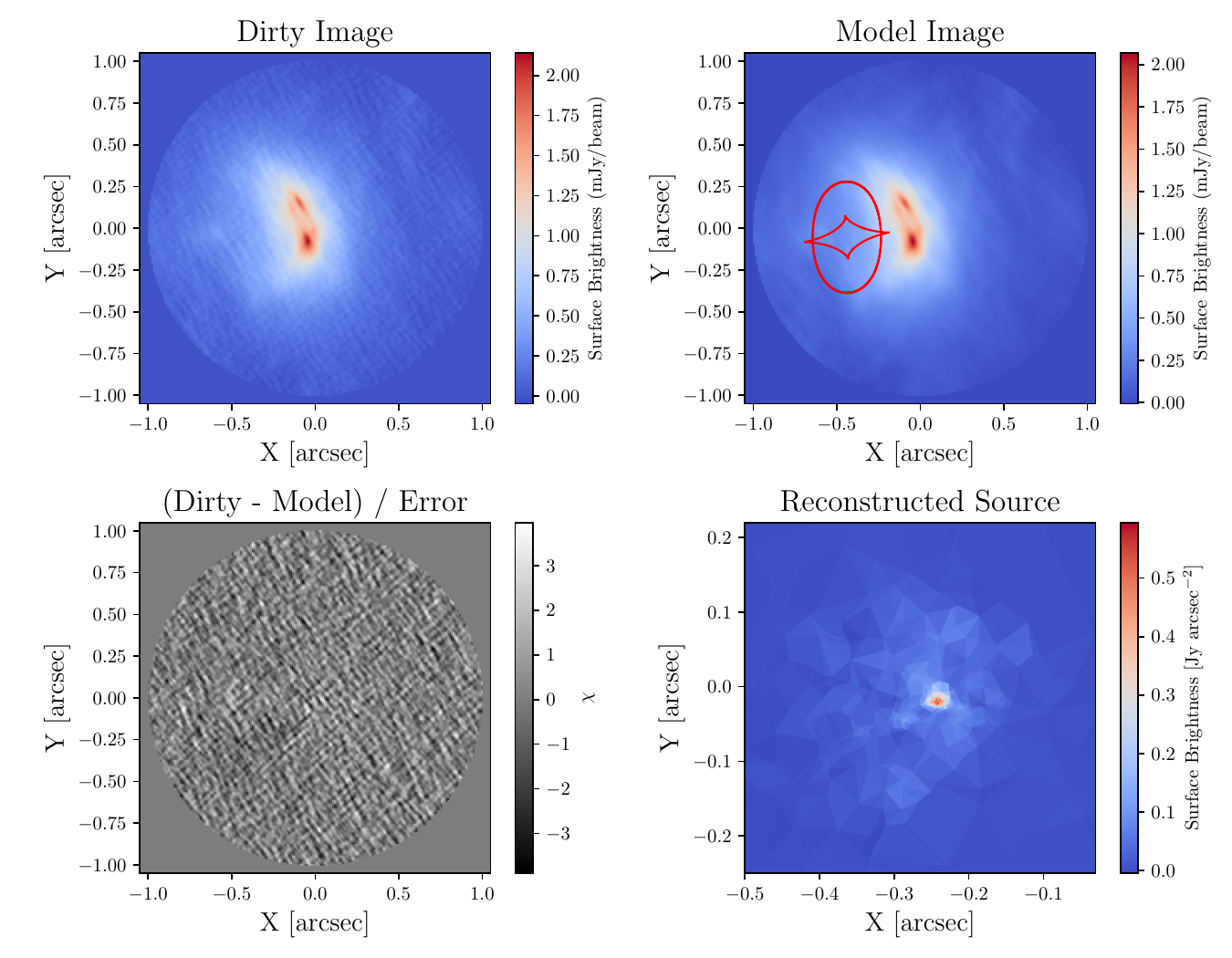}
    \caption{The best-fit lensing model for the dust continuum. {\em Top left:} the observed dirty image; {\em Top right:} the best-fit model dirty image; {\em Bottom left:} the residual map; {\em Bottom right:} the source-plane emission. The red lines mark the caustics of the lensing potential, which exhibits a naked-cusp structure. The source plane emission shows a compact bright core and a fainter outskirt with some irregular features. The total flux is {$3.36\pm0.02$ mJy}, and the flux magnification is {$4.63\pm0.03$}. The source-plane spatial resolution is $0\farcs019$ (about 104 pc).}
    \label{fig:lensmodel}
\end{figure*}

Figure \ref{fig:lensmodel} shows the best-fit model and the corresponding source plane emission. The residual dirty image is dominated by random noise and does not show any significant pattern, indicating that the model we choose is adequate. 
The lensing caustics exhibit a naked-cusp structure, i.e., the tangential caustics (the red diamond-shaped curve) extend outside of the radial caustics (the red elliptical curve). This result is qualitatively consistent with the lensing model in \cite{fan19}. 
The source-plane emission shows a compact core $(r\lesssim0\farcs03)$ with a high peak surface brightness $(0.6 \text{ Jy arcsec}^{-2})$, indicating significant AGN heating in this region. The outskirts $(r\gtrsim200 \text{ pc})$ show a disk-like structure with some irregular features. 

The total flux of the source-plane emission is {$F_{245{\rm GHz}}=3.36\pm0.02$ mJy }, where the uncertainty is derived using the covariance matrix $\boldsymbol{\Sigma}_s$. This flux is slightly lower than the value reported by \citet{yue21}, who used a {S\'ersic} profile to fit the quasar host galaxy and found $F_{245{\rm GHz}}=3.46 \pm 0.04$ mJy. The best-fit lensing model indicates a total flux magnification of {$4.63$}; correspondingly, we estimate the average source-plane spatial resolution to be {$0\farcs04/\sqrt{4.63}=0\farcs019$}, i.e., about 104 pc at $z=6.52$. We note that the spatial resolution varies across the source plane; generally speaking, regions closer to the tangential caustics (the diamond-shaped red curve in Figure \ref{fig:lensmodel}) have higher magnification and thus finer spatial resolution.


Previous low-resolution ALMA observations for high-redshift quasars \citep[e.g.,][]{decarli18,venemans20} usually use Gaussian profiles or exponential disks to fit their host galaxy emissions. Nevertheless, these analytical profiles cannot well-describe the bright compact core at the center. To evaluate the size of the host galaxy, we fit an exponential disk to the region with $0\farcs03<r<0\farcs3$ centered at the brightest pixel, mitigating the impact on the central bright core. We
use the covariance matrix of the source plane pixels ${\bf\Sigma}_{\rm s}$ when fitting the profile.  This fitting yields a half-light radius of {$r_e=0\farcs148\pm0\farcs003$ (about 0.8 kpc), with an axis ratio of $q=0.76^{+0.02}_{-0.01}$ and a position angle of $\omega=68.96^{+1.65}_{-1.44}$ degrees}. The half-light radius is similar to other $z>6$ quasars, and close to the results in \citet{yue21} who suggested $r_e=0\farcs136$ for J0439+1634 using low-resolution ALMA observations.

We also compare the FIR continuum map of J0439+1634 to other $z>6$ quasars with $<500$ pc resolution ALMA imaging. These targets includes J0305–3150 \citep[][]{venemans19}, J2348–3054 \citep{walter22}, and J0109–3047 \citep[][]{meyer23}. Interestingly, J0305–3150 exhibits similar features to J0439+1634, including a compact bright core in its FIR map, and an extended, slightly irregular disk. In contrast, J2348–3054 and J0109–3047 have smaller sizes in FIR emission and do not show prominent extended disks. These comparisons demonstrate the diversity of $z>6$ quasar host galaxy morphologies.

\subsection{Re-analyzing HST Observations} \label{sec:newhst}


In the discovery paper, \citet{fan19} presented the {\em HST}-based lensing model for J0439+1634, which was later used in \citet{yue21} to analyze the C43-5 ALMA observations. When building the lensing model, \citet{fan19} used a single SIE model without external shears, where the position and shape of the SIE were fixed to the foreground galaxy emission in the {\em HST} images.
The {\em HST} observations have an angular resolution of $\sim0\farcs1$, which is several times larger than the ALMA observation presented in this work.
We therefore prefer the ALMA-based lensing model to the {\em HST-}based one, and re-analyze the {\em HST} image using the ALMA-based lensing model.

Specifically, we obtain the {\em HST} Advanced Camera for Survey Wide Field Camera (ACS/WFC) F782N filter image and the point spread function (PSF) from \citet{fan19}. We use \texttt{glafic} \citep[][]{glafic} to fit the F782N image, where we use a point source to describe the quasar emission. The position and flux of the point source are left free to change. We fix the lensing potential parameters to the best-fit ones in the ALMA-based model, except for the position of the SIE. We use \texttt{glafic} task \texttt{optimize} to find the best-fit model.

\begin{figure*}
    \centering
    \includegraphics[width=0.7\linewidth]{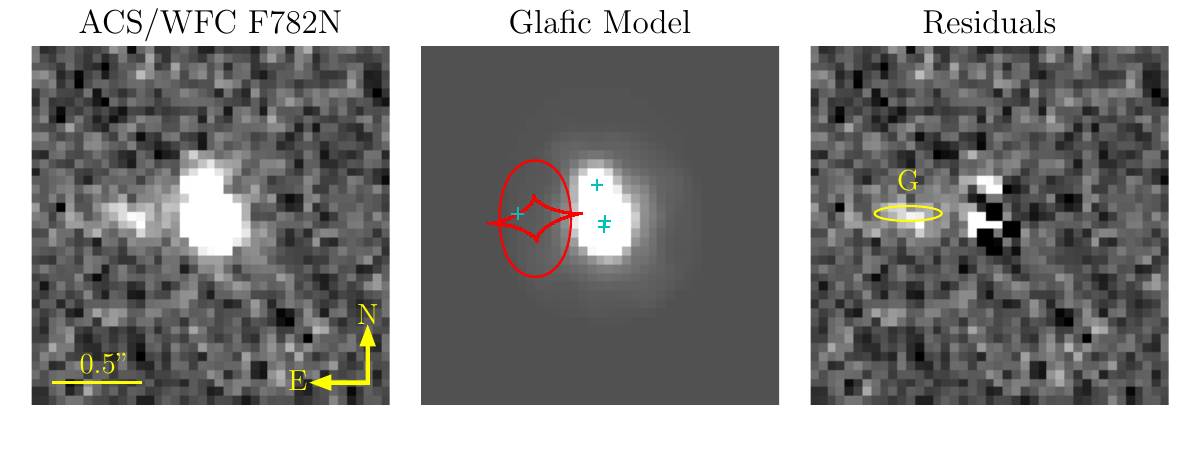}
    \caption{Re-analyzing the {\em HST} image using the ALMA-based lensing model. {\em Left:} the ACS/WFC F782N image, which was used to build the lensing model in \citet{fan19}. {\em Middle:} the best-fit image using the ALMA-based lensing model. The red lines mark the lensing caustics, and the cyan crosses mark the lensed images of the optical quasar. {\em Right:} the residual image. The yellow ellipse shows the position and shape (i.e., position angle and ellipticity) of the SIE, which agrees well with the foreground galaxy emission in the {\em HST} image.  The updated model indicates a total magnification of 37.1 for the optical quasar.}
    \label{fig:hst}
\end{figure*}

Figure \ref{fig:hst} presents the best-fit model for the {\em HST} image. 
Notably, the position and shape of the SIE component in the ALMA-based lensing model agree well with the foreground galaxy emission in the {\em HST} image (Figure \ref{fig:hst}, right panel). 
We also compute the relative position between the optical quasar and the SIE, and find that the optical quasar is only $0.8$ milli-arcsecond (mas) away from the brightest pixel in the sub-mm continuum.
These comparisons confirm that the lensing model in Section \ref{sec:lensmodeling} correctly describes the lensing configuration of J0439+1634.

Unlike the {\em HST-}based lensing model where the optical quasar has three lensed images, the updated model suggests four lensed images for the quasar. The fourth image, located at the east side of the lensing galaxy, is highly demagnified (with $\mu=0.01$). The other three lensed images have magnifications of $3.8$, $16.4$, and $16.9$, respectively, yielding a total magnification of $37.1$. Accordingly, we estimate the source-plane resolution around the SMBH to be $\sim0\farcs04/\sqrt{37.1}=6.56$ mas, equivalent to $36$ pc at $z=6.52$. Therefore, the ``bright spot" on the reconstructed source plane map is resolved by our observation.

\section{AGN Contribution to Dust Heating in the Quasar Host Galaxy} \label{sec:heating}

\subsection{Radiative Transfer Model}
The source plane sub-mm emission has a bright peak with a surface brightness of $0.6\text{ Jy arcsec}^{-2}$. If we assume that star formation dominates the dust heating at the galaxy center, the peak brightness would imply an SFR surface density of $9\times10^3M_\odot\text{ yr}^{-1}\text{kpc}^{-2}$, following the analysis in \cite{yang19} and \cite{yue21}. This value is nearly 1 dex higher than the highest SFR surface density observed in the Universe \citep[$\sim10^3 M_\odot\text{ yr}^{-1}\text{kpc}^{-2}$; e.g.,][]{walter09}. Therefore, AGN-heated dust must dominate sub-mm emission at the center of the quasar host galaxy.

With the help of lensing magnification, we can investigate the quasar's contribution to its host galaxies' dust heating at different scales. To do this, we build a toy model using the radiative transfer simulation software \textsc{skirt} \citep[][]{skirt}. \textsc{skirt} performs Monte Carlo simulations on 3D grids and models dust scattering, absorption, and emission processes.
The toy model consists of the following components:

- A central AGN. We use the default quasar spectral energy distribution (SED) in \textsc{skirt}, which is described by a broken power-law (see the document\footnote{https://skirt.ugent.be/skirt9/class\_quasar\_s\_e\_d.html} for more details). \cite{fan19} measured the apparent  3000{\AA} luminosity of J0439+1634 to be $4.35\times10^{47}\text{ erg s}^{-1}$. Since the optical quasar has a magnification of $\mu=37.1$, we set the 3000{\AA} luminosity of the AGN component to be $L_{3000}=1.17\times10^{46}\text{erg s}^{-1}$.

- A dust torus around the AGN. We assume a power-law density distribution, i.e., $\rho\propto r^{-1}$, in the volume $1.4\text{ pc}<r<700 \text{ pc}$. This radius range is similar to previous modeling for AGN dust structures \citep[e.g.,][]{lyu18,lyu21}. We set the opening angle to be $30^\circ$. \cite{yang22} measured the hydrogen column density of J0439+1634 to be ${N}_{{\rm{H}}}={2.8}_{-0.6}^{+0.7}\times {10}^{23}\,{\mathrm{cm}}^{-2}$, using X-ray observations by XMM-Newton. Correspondingly, we set the $V-$band optical depth of the torus to be $A_V=100$, given the typical $N_{\rm{H}}/A_V$ ratio of $\sim2\times10^{21}{\rm cm}^{-2}{\rm mag}^{-1}$ \citep[e.g.,][]{tolga09}. 

{We note that the specific parameters for the dust torus component do not have a significant impact on this experiment. We demonstrate this by removing the torus component in the fitting process. The difference between the no-torus best-fit model and the fiducial best-fit model is about $2\%$ for their $\chi^2$, and the inferred host galaxy parameters remain the same (see below). Both the no-torus fit and the fiducial fit indicate that AGN heating contributes $<20\%$ of the total FIR flux (see Section \ref{sec:implication} for details).}



- The host galaxy. We assume an exponential thin disk for the host galaxy, characterized by the scale radius $R_\text{exp}$. We assume that the star and dust contents in the host galaxy have the same spatial distribution. The dust content is normalized by the total dust mass, $M_\text{dust}$. The star component is normalized by the total UV luminosity, $L_{1500}$. We use a single-age stellar population to describe the star component, with a metallicity of $\log (Z/Z_\odot)=-0.5$ and a stellar age of $0.6$ Gyrs. These numbers are similar to recent James Webb Space Telescope measurements for high-redshift quasars \citep[e.g.,][]{marshall25}. To find the parameters that best match the observed emission, we allow $R_\text{exp}$, $M_\text{dust}$, and $L_{1500}$ to vary on a grid:

\begin{itemize}
    \item $R_\text{exp}$ varies on a linear grid at $[300 {\text{ pc}}, 1000 {\text{ pc}}]$ with an increment of 50 pc;
    \item $\log M_\text{dust} ~[M_{\odot}]$ varies on a linear grid at [8.7, 9.5] with an increment of 0.08 dex;
    \item $\log L_\text{1500} \text{ [erg s}^{-1}]$ varies on a linear grid at [44.5, 45.5] with an increment of 0.1 dex.
\end{itemize}

To generate mock observations,
\textsc{skirt} also takes the inclination $(\theta)$ and rotation angle $(\omega)$ of the observing instrument as inputs. Fitting these parameters requires running a large number of simulations and is computationally expensive. Therefore, we determine $\theta$ and $\omega$ using the host galaxy fitting described in Section \ref{sec:analysis}, where we fit an exponential profile to the annuli with $0\farcs03<r<0\farcs3$. {The adopted values are $\theta=\arccos(q)=40.54^\circ$ and $\omega=68.92^\circ$, respectively.}

With the $\theta$ and $\omega$ values determined above, we run \textsc{skirt} simulation for the $R_\text{exp}-M_\text{dust}-L_{1500}$ grid. We use the source plane covariance matrix $\boldsymbol{\Sigma}_s$ to determine the best-fit model. {We find that $R_\text{exp}=550$ pc, $\log M_\text{dust}[M_\odot]=9.1$, and $\log L_{1500} [{\rm erg~s}^{-1}]=45.1$ match the observation best.} 

\begin{figure*}
    \centering
    \includegraphics[width=1\linewidth]{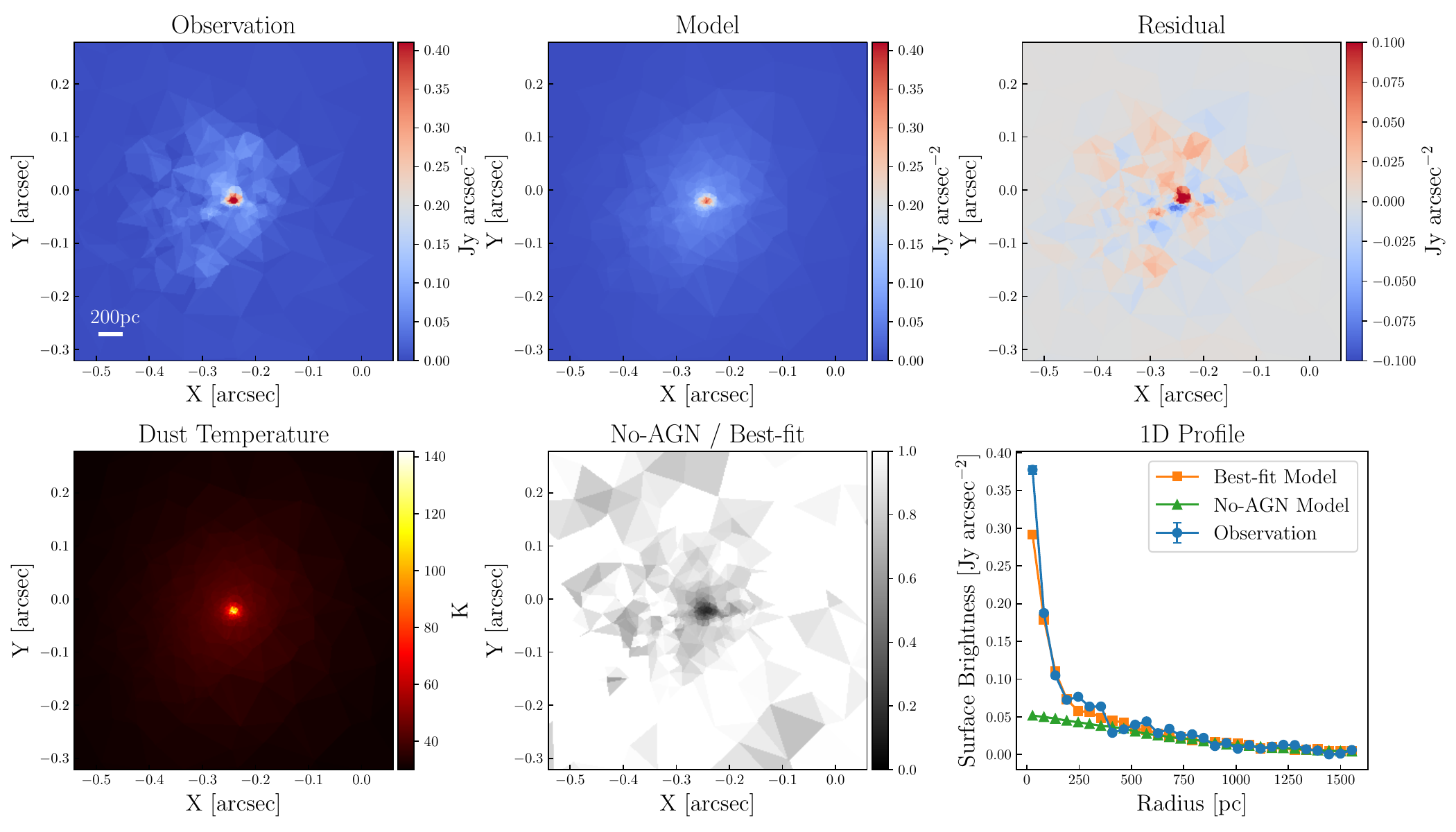}
    \caption{Modeling the sub-mm continuum emission of J0439+1634 using \textsc{skirt}. {\em Top left:} the {reconstructed} source-plane surface brightness; {\em Top middle:} the best-fit \textsc{skirt} model; {\em Top right}: the residual; {\em Bottom left}: the dust temperature map of the best-fit model; {\em Bottom middle}: the flux ratio between the no-AGN model and the best-fit model; {\em Bottom right}: the 1D profile of the {source-plane} surface brightness, the best-fit model, and the no-AGN model, where the X-axis is the semi-major axes of the elliptical annuli (see text). AGN-heated dust dominates the sub-mm flux at $r\lesssim100$ pc, while the outer region is dominated by star-heated dust.}
    \label{fig:skirt}
\end{figure*}

Figure \ref{fig:skirt} shows the best-fit \textsc{skirt} model.
Despite its simplicity, the toy model provides a decent description of the observation. We also compute the 1D profile of the surface brightness using a series of elliptical annuli, {which have an axis ratio of $q=0.76$ and rotation angles of $\omega=68.92^\circ$,} i.e., same as the host galaxy. The lower right panel of Figure \ref{fig:skirt} shows the 1D profile. {The \textsc{skirt} model is close to the reconstructed source profile at almost all radii, except for the innermost bin $(r<0\farcs01)$ where the model underpredicts the surface brightness. This difference may reflect a deviation of the central dust distribution from an exponential disk profile (see \ref{sec:limitation} for more discussion). We note that the innermost bin only contributes to about $3\%$ of the total flux.}


\subsection{Implications for Quasar Host Galaxy Studies} \label{sec:implication}

Previous studies \citep[e.g.,][]{decarli18,yang19,venemans20,wang24} have used sub-millimeter flux to evaluate the FIR luminosities of high-redshift quasars and derive their SFRs. These studies usually assume a dust temperature of $T_d=47$K and that the sub-mm emission is dominated by dust heated by stellar emission. 
Meanwhile, \citet{tsukui23} suggested that AGN-heated dust might has a significant contribution to the FIR flux of quasar host galaxies, and that the center region $(r\lesssim1 \text{kpc})$ can have dust temperatures $T_d\gtrsim50$ K. Similarly, \citet{dmf23} suggested that the TIR-based SFR of $z\sim6$ quasars might be overestimated by over an order of magnitude. Therefore, it is important to test the above-memtioned assumptions about FIR luminousities and SFRs of high-redshift quasars.


With the radiative transfer model for J0439+1634, we can evaluate how AGN-heated dust biases the measurements of FIR luminosity and SFR. We first evaluate the FIR flux without quasar activity (i.e., only driven by stellar emission). We do this by removing the AGN component and re-run the best-fit \textsc{skirt} model. {The no-AGN model yields a total flux of 3.10 mJy, compared to 3.56 mJy for the best-fit \textsc{skirt} model.
In other words, although AGN heating produces high surface brightness at the center, star-heated dust still contributes $\sim87\%$ total sub-mm flux of the quasar host galaxy.} This finding is similar to that of \citet{silverman26}, who analyzed three $z\sim2$ luminous quasars and found that AGN contributes $\lesssim10\%$ to their observed FIR fluxes.

 We also examine whether the dust temperature assumption is valid. In the bottom left panel of Figure \ref{fig:skirt}, we show the dust temperature map of the best-fit model. The outer region $(r\gtrsim100$ pc) of the quasar host have $T_d\sim40{\rm K}-60{\rm K}$, for which $T_d=47$K is a reasonable assumption. 

We further investigate the spatial extent over which AGN dominates dust heating. The bottom middle panel of Figure \ref{fig:skirt} shows the ratio between the no-AGN model and the best-fit model, and the bottom right panel shows the 1-D profile of the on-AGN (i.e., host-only) model. At $r\lesssim100$ pc, the ratio is below $50\%$, indicating that AGN-heated dust dominates the sub-mm emission in this region. The host galaxy contribution dominates the flux at $r\gtrsim200$ pc.


\subsection{{Limitation of the model}} \label{sec:limitation}

{
The toy model described in this Section is oversimplified in many aspects.
One crucial limitation is that we use an exponential thin disk to describe the host galaxy morphology and do not consider any irregularities. Indeed, galaxies might have compact star-forming regions in their central 100 pc \citep[e.g.,][]{fm25}. 
Although the high surface brightness peak seen in Figure \ref{fig:lensmodel} exceeds the Eddington limit of SFR surface densities \citep[see discussions in ][]{yue21}, recent simulation studies have found that the SFR surface densities can reach $\sim10^4{M_\odot}{\rm yr^{-1}kpc^{-2}}$ in the central $\sim10$ pc of galaxies \citep[e.g.,][]{aa21}. In the case of J0439+1634, the luminous optical quasar already indicates the presence of AGN heating, and the central $\sim10$ pc is unresolved even with lensing magnification. Therefore, we still interpret the bright core as a sign of AGN heating, with a note that  compact star-forming regions might also have non-negligible contributions.}

{
We also note other limitations of our model, including (1) we assume that the dust and stellar components have the same spatial distribution; (2) we assume a single-age stellar population for the host galaxy SED; and (3) we do not consider different dust grain properties and use the default \textsc{skirt} dust mixture model.}

With all these limitations, we argue that the ``best-fit" parameters in our model (i.e., $M_\text{dust}$, $L_\text{UV}$, and $R_\text{exp}$) should be taken as rough estimates. Future higher-frequency ALMA observations will measure the dust temperature map for J0439+1634 more accurately, putting stronger constraints on the AGN's contribution to dust heating in the quasar host galaxy.

\section{Conclusion} \label{sec:conclusion}

We present high-resolution ALMA observations of J0439+1634, a gravitationally-lensed quasar at $z=6.52$. The image-plane resolution is $0\farcs078\times0\farcs040$. We use \texttt{PyAutoLens} to constrain the lensing potential and the source-plane sub-mm continuum emission. 
Using the reconstructed source plane emission, we build a radiative transfer model to evaluate the contribution of AGN-heated dust to the sub-mm flux. Our main findings include

\begin{enumerate}
    \item The lensing potential of J0439+1634 has a naked-cusp configuration, similar to the lensing model in \cite{fan19}.  The quasar host galaxy has a compact, bright core with a peak brightness of $0.6\text{ Jy arcsec}^{-2}$. 
    {The quasar host galaxy has a total flux of $F_\text{245GHz}=3.36\pm0.02$ mJy, and the overall magnification is $4.63\pm0.03$.} We estimate the average source-plane angular resolution to be $0\farcs019$ (about 104 pc).
    
    \item By re-fitting the {\em HST} image, we find that the optical quasar has four lensed images, one of which is highly demagnified. The total magnification of the optical quasar is 37.1, implying a source-plane resolution of 36 pc near the SMBH. The position of the optical quasar is consistent with the brightest pixel in the source-plane sub-mm continuum emission.
    \item The AGN dominates dust heating in the inner region $(r\lesssim100\text{ pc})$ of the quasar host galaxy. Meanwhile, star-heated dust dominates the outer region of the quasar host galaxy and contributes $\sim80\%$ of the total sub-mm continuum flux. We suggest that the FIR-based SFR estimator mildly overestimates the SFR {(by $\sim13\%$)} of high-redshift quasars like J0439+1634.
\end{enumerate}

J0439+1634 demonstrates the power of lensing magnification in obtaining deep insights into distant SMBHs and their host galaxies. The new-generation facilities, like the {\em Euclid} telescope, the Rubin Observatory Legacy Survey for Space and Time, and the Nancy Grace Roman Space Telescope, are expected to yield a much larger sample of strongly lensed AGNs at $z\gtrsim6$ \citep[e.g.,][]{yue22a, yue22b}. This {forthcoming} sample will enable deep investigations of SMBHs and their coevolution with galaxies in the early Universe.

The code and data used in this work will be available at 10.5281/zenodo.20320289.

\begin{acknowledgments}
We thank the referee for the valuable and constructive comments, which have helped us to greatly improve the manuscript.
MY acknowledges support from the Bart J. Bok Fellowship from the University of Arizona.
AIZ acknowledges support from NASA ADAP
grant 80NSSC21K0988.

\end{acknowledgments}





%
\facilities{ALMA}

\software{astropy \citep{2013A&A...558A..33A,2018AJ....156..123A,2022ApJ...935..167A},  
            CASA \citep[][]{casa},
          PyAutoLens, 
          SKIRT
          }



\bibliography{sample701}{}
\bibliographystyle{aasjournalv7}



\end{document}